\documentclass[a4paper,twocolumn,11pt,allowtoday]{quantumarticle}
\pdfoutput=1
\usepackage[utf8]{inputenc}
\usepackage[english]{babel}

\usepackage[T1]{fontenc}
\usepackage{microtype}   
\usepackage{bm}

\usepackage{mathtools}   
\usepackage{amssymb}
\usepackage{braket}
\usepackage{mathrsfs}    

\usepackage{graphicx}
\usepackage[table]{xcolor}
\usepackage{array}
\usepackage{booktabs}     
\usepackage{multirow}
\usepackage{diagbox}
\usepackage{dcolumn}      
\usepackage{makecell} 
\usepackage{siunitx}

\usepackage{float}
\usepackage{indentfirst}
\usepackage[normalem]{ulem} 
\usepackage{ragged2e}
\usepackage{enumitem}     

\usepackage{algorithm}      
\usepackage{algpseudocode}  

\usepackage[numbers,sort&compress]{natbib}
\usepackage{hyperref}

\newcommand{\OU}{Graduate School of Engineering Science, The University of Osaka, 1-3 Machikaneyama, Toyonaka, Osaka 560-8531, Japan}
\newcommand{\FI}{Center for Computational Quantum Physics, Flatiron Institute, New York, NY 10010, USA}

\newcommand{\QIQB}{Center for Quantum Information and Quantum Biology, Osaka University, Toyonaka, 560-0043, Japan}

\newcommand{\RCCS}{RIKEN Center for Computational Science (R-CCS), Kobe, 650-0047, Japan.}

\begin{document}

\title{Quantum-Inspired Algorithm for Classical Spin Hamiltonians Based on Matrix Product Operators}

\author{Ryo Watanabe}
\email{ryowatanabe06jp@gmail.com}
\affiliation{\OU}

\author{Joseph Tindall}
\affiliation{\FI}

\author{Shohei Miyakoshi}
\affiliation{\QIQB}
\affiliation{\RCCS}

\author{Hiroshi Ueda}
\affiliation{\QIQB}
\affiliation{\RCCS}

\begin{abstract}
We propose a tensor-network (TN) approach for solving classical optimization problems that is inspired by spectral filtering and sampling on quantum states.
We first shift and scale an Ising Hamiltonian of the cost function so that all eigenvalues become non-negative and the ground states correspond to the the largest eigenvalues, which are then amplified by power iteration.
We represent the transformed Hamiltonian as a matrix product operator (MPO) and form an immense power of this object via truncated MPO-MPO contractions, embedding the resulting operator into a matrix product state for sampling in the computational basis.
In contrast to the density-matrix renormalization group, our approach provides a straightforward route to systematic improvement by increasing the bond dimension and is better at avoiding local minima.
We also study the performance of this power method in the context of a higher-order Ising Hamiltonian on a heavy-hexagonal lattice, making a comparison with simulated annealing.
These results highlight the potential of quantum-inspired algorithms for solving optimization problems and provide a baseline for assessing and developing quantum algorithms.
\end{abstract}

\maketitle

\section{Introduction}\label{sec:Introduction}

Classical spin-glass Hamiltonians not only play a central role in the theoretical physics of disordered and interacting systems~\cite{S.F.Edwards_1975,RevModPhys.58.801}, 
but also provide a framework for formulating various combinatorial optimization problems (COPs) 
that arise in real-world applications~\cite{Applegate2007TSP,TothVigo2002,Burkard2012Assignment,Pinedo2022,Melo2009FacilityLocation,FordFulkerson1956,KorteVygen2018}.
These Hamiltonians exhibit frustration in an exponentially large configuration space, originating from inharmonious or competing interactions.
Therefore, finding their
ground states is generally NP-hard, as in the case for many COPs~\cite{Lucas2014, garey2002computers}. Unless $\mathrm{P}=\mathrm{NP}$, no polynomial-time algorithm is known to solve them exactly.

Many classical algorithms employ heuristics and approximations, such as simulated annealing (SA)~\cite{SA} and genetic algorithms (GA)~\cite{anderson1994genetic}, to tackle large-scale instances, often yielding near-optimal solutions.
Meanwhile, quantum hardware has recently been considered a promising approach because it can directly manipulate wavefunctions in Hilbert space.
Prominent paradigms include quantum imaginary time evolution~\cite{Alam2023,PhysRevA.109.052430}, quantum adiabatic evolution~\cite{Farhi_2001}, quantum annealing (QA)~\cite{finnila1994quantum, doi:10.1126/science.ado6285}, and the quantum approximate optimization algorithm (QAOA)~\cite{farhi2014quantumapproximateoptimizationalgorithm}.
These quantum algorithms for solving COPs have also, in turn, motivated a number of ``quantum-inspired approaches" to solving these problems on classical computers.

In these quantum-inspired approaches, tensor-networks (TNs) are frequently employed \cite{morais2025comparative, hao2022quantum, nakada2025quick} due to their efficiency in representing low-entanglement quantum states~\cite{orus2014practical,orus2019tensor,Berezutskii2025}.
The well-established density-matrix renormalization group (DMRG) algorithm~\cite{Schollw_ck_2011}, for instance,
can be directly applied \cite{morais2025comparative} to a diagonal, Matrix Product Operator (MPO) representation of the COP cost function $\hat{H}$ in an attempt to find the Matrix Product State (MPS) --- a one dimensional TN --- which minimizes $\bra{\psi} \hat{H} \ket{\psi}$.
This approach is, however, prone to becoming trapped in local minima when complex energy landscapes emerge with large degeneracies --- as is frequently the case in classical, frustrated, energy landscapes.
Careful choice of hyperparameters in the DMRG algorithm, such as the initial state, learning rate, and additional noise, can be utilised to try to alleviate such problems~\cite{PhysRevB.72.180403}.

In this study, we propose an alternate, tensor network-based approach for solving COPs based on spectral amplification and filtering of a positive semi-definite MPO representation of the cost function.
Specifically, we consider the shifted and scaled Hamiltonian $\hat{G}=a\,\hat{H}+b$ with $a,b\in\mathbb{R}$, choosing $a$ and $b$ so that $\hat G$ is positive semidefinite and the eigenvalues corresponding to the ground states of $\hat{H}$ are the maximum eigenvalues of $\hat{G}$.
We then represent $\hat{G}$ as an MPO with a bond dimension that can be tightly bounded in terms of the properties of the underlying cost function. Next, we perform repeated MPO-MPO multiplication to construct $\hat{G}^K$ with $K \gg 1$, massively amplifying the lowest energy configurations of $\hat{H}$.
This powered MPO is diagonal in the computational basis and thus can be immediately transformed into an MPS, allowing perfect sampling in the computational basis.
Through this procedure, one can then extract high quality candidate solutions to the original optimization problem.

While our approach shares the spectral-filtering philosophy of imaginary time evolution, it remains entirely Trotter-free.
Moreover, compared to quantum imaginary time methods that rely on gate decompositions and parameterized unitaries~\cite{ Alam2023,PhysRevA.109.052430} to operate within the paradigm of quantum computing, our formulation embeds the non-unitary cost function directly into a single classical MPO upon which spectral filtering can be efficiently performed via MPO-MPO contraction. Furthermore, our approach naturally facilitates exponentially fast spectral filtering by repeatedly multiplying the MPO by itself.

We benchmark our approach on both Edwards--Anderson spin glass model on hypercubic lattices~\cite{S.F.Edwards_1975,Boettcher_2024,Zhang_2025} and higher-order Ising spin-glass models defined on the heavy-hexagonal lattice which has become a standard benchmark for QAOA on current IBM quantum devices~\cite{QA_vs_QAOA_127, Short_Depth_QAOA_QA}.
We observed that the powered-MPO approach reliably identifies low-energy configurations on these problems even when hundreds of variables are involved.

Compared with DMRG, we find our method provides a more robust mechanism for escaping local minima.
While DMRG's sweeps are highly cost-effective at finding near-optimal states, they are susceptible to stagnation in the cost landscape. In contrast, the powered-MPO method exhibits more systematic improvement with the bond dimension $\chi$.
Furthermore, this superiority extends to comparisons with SA, a widely used heuristic for combinatorial optimization.
In challenging instances where SA remains trapped in suboptimal local minima, our powered-MPO approach consistently identify superior answers.

The remainder of this paper is organized as follows.
Section \ref{sec:preliminaries} introduces the operator basis for the cost Hamiltonian and provides a theoretical analysis of the power iteration.
The shifting and scaling strategy required to render the cost Hamiltonian positive semidefinite is formulated precisely in Appendix \ref{appendix:shifted_scaled_Hamiltonian}.
This Hamiltonian can be represented as an MPO as described in Appendix~\ref{appendix:mpo_construction}.
Section \ref{sec:method} describes the procedure for building powers of the MPO and embedding it into an MPS to sample from it in the computational basis.
In Section \ref{sec:benchmark}, we demonstrate the performance of the proposed method through numerical experiments on a variety of spin-glass models.
Finally, Section \ref{sec:conclusion} summarizes our findings and discusses possible directions for future research.

\section{Preliminaries}\label{sec:preliminaries}

\subsection{Operator Basis for classical Spin Hamiltonians}\label{subsec:operator_basis}

We first consider cost functions $C(\bm{z}) \in \mathbb{R}$ over heterogeneous $d_i$-ary strings $\bm{z}=(z_1,\ldots,z_N)$ with $z_i\in\{0,\ldots,d_i-1\}$. 
We encode this function as a Hamiltonian $\hat H$ that is diagonal in the standard product basis (computational basis) $\mathcal{B}=\{\ket{\bm{z}}=\bigotimes_{i=1}^N \ket{z_i}\}$, namely
\begin{equation}
\label{eq:costhamiltonian}
\hat H \;=\; \sum_{\bm{z}} C(\bm{z})~\ket{\bm{z}}\bra{\bm{z}}~,
\end{equation}
acting on the real Hilbert space $\mathcal H=\bigotimes_{i=1}^N \mathcal H_i$ with $\mathcal H_i\cong\mathbb{R}^{d_i}$, whose total dimension is $\mathcal{D} = \prod^N_{i=1} d_i$.
Additionally, we assume that $\hat H$ admits an expansion in terms of $N_P$ linearly independent operators on $\mathcal H$ with unique coefficients.
With this setup in place, we aim to find the ground state of $\hat{H}$, i.e the classical configuration(s) that minimize(s) $C(\bm {z})$.

Let us define an operator basis $\mathcal{O}$ that is orthonormal with respect to the Hilbert–Schmidt inner product.
This basis allows us to expand Eq.~\eqref{eq:costhamiltonian} as 
\begin{equation}
\label{eq:costhamiltonian_op}
\hat H = \sum_{j=1}^{N_P} h_j \hat{O}_j~,
\end{equation}
by using a subset $\{\hat O_j\}_{j=1}^{N_P}\subset\mathcal O$ and $h_j \in \mathbb{R}$, and these coefficients $\{ h_j \}^{N_P}_{j=1}$ are uniquely determined.

Recall that we work with general qudits in the computational basis $\mathcal{B}$, where $z_i \in \{ 0,\ldots,d_i-1 \}$.
At each site $i$, we can define
\begin{equation}\label{eq:local_operator_basis}
  \mathcal P_i := \{ \hat{Z}^{(0)}_{i}, \hat{Z}^{(1)}_{i}, \ldots, \hat{Z}^{(d_i-1)}_{i} \},
\end{equation}
where $\hat{Z}^{(0)}_{i} := \hat I_i$ is the identity operator on the local Hilbert space $\mathcal{H}_i$,
and $\{ \hat Z^{(k)}_{i} \}_{k=1}^{d_i-1}$ are chosen as traceless, Hermitian, diagonal generators of $\mathrm{SU}(d_i)$ that span its Cartan subalgebra.
Explicitly, for $k>0$,
\begin{equation}
    \hat{Z}^{(k)}_i = \sqrt{\frac{d_i}{k(k+1)}}
    \operatorname{diag}(\overbrace{1,\ldots,1}^{k}, -k, \overbrace{0,\ldots,0}^{d_i-k-1})~,
\end{equation}
so that $\mathcal P_i$ forms an orthogonal basis with respect to the Hilbert–Schmidt inner product:
\begin{equation}\label{eq:orthogonality}
\begin{aligned}
\mathrm{Tr}(\hat Z^{(k)}_{i} \hat Z^{(\ell)}_{i}) &= d_i  \delta_{k,\ell}~, \\
\mathrm{Tr}(\hat Z^{(k)}_{i}) &= 
\begin{cases}
d_i & \text{if } k=0~, \\
0   & \text{if } k>0~,
\end{cases}
\end{aligned}
\end{equation}
where $\delta_{k,\ell}$ is the Kronecker delta~\cite{georgi2000lie}.
With these local operator sets, the global operator basis is defined as
\begin{equation}\label{eq:operator_basis}
  \mathcal O := \Big\{ \bigotimes_{i=1}^N \hat{P}_i \;\Big|\; \hat{P}_i \in \mathcal P_i \Big\}~.
\end{equation}

In the particular case where $d_i = 2$ for all sites, we take $\mathcal{O} = \{\hat I, \hat Z\}^{\otimes N}$, with $\hat Z$ denoting the Pauli-$Z$ operator.
This setting defines the classical Ising Hamiltonian. The Ising formulation provides a standard representation for many COPs over binary variables, which can be formulated either i) as a quadratic unconstrained binary optimization (QUBO) problem, where the Hamiltonian includes linear and quadratic interactions, or ii) more generally as a higher-order unconstrained binary optimization (HUBO) problem that allows multi-body interaction terms~\cite{Kochenberger2014,Lucas2014}.

\subsection{Theoretical Analysis of the Powered-Operators}\label{subsec:theoretical_analysis}

Since our goal is to extract the ground states of $\hat{H}$, we transform the cost Hamiltonian $\hat{H}$ as
\begin{equation}\label{eq:G}
\hat{G} = a\hat{H} + b\mathbb{I}~,
\end{equation}
where $\mathbb{I}$ represents the identity operator, and $a,b \in \mathbb{R}$ are chosen such that the target eigenvalue of $\hat H$ corresponds to the spectral radius of $\hat G$, i.e., the largest eigenvalue by magnitude. We further tune $a$ and $b$ to impose positive semidefiniteness on $\hat{G}$ 
and prevent the simultaneous amplification of both maximal and minimal eigenvalues in $\hat{G}^K$.

We find the positive semidefiniteness of $\hat{G}$ helps to suppress unnecessary growth of the bipartite entanglement entropy (EE) in the power of $\hat G^K$.
Formally, since $\hat G^K$ is diagonal in the computational basis, it can be embedded into the pure state
\begin{equation}\label{eq:Psi}
  \ket{\Psi} \propto \hat{G}^K \bigotimes_{i=1}^N \ket{+}_i~,
\end{equation}
where $\ket{+}_i$ is a state defined as
\begin{equation}\label{eq:uniform_state}
    \ket{+}_i = \frac{1}{\sqrt{d_i}} \sum^{d_i-1}_{z_i=0} \ket{z_i}~.
\end{equation}
In this construction, the probability of each configuration is proportional to the square of the corresponding diagonal element of $\hat G^K$.
Accordingly, the bipartite EE of $\ket{\Psi}$ is defined as follows:
\begin{equation}\label{eq:EE_def}
  \mathcal{S} = - \mathrm{Tr}_A(\rho_A \log_2 \rho_A) 
              = - \mathrm{Tr}_B(\rho_B \log_2 \rho_B)~,
\end{equation}
with $\rho_A = \mathrm{Tr}_B({\ket{\Psi}\bra{\Psi}})$ and 
$\rho_B = \mathrm{Tr}_A({\ket{\Psi}\bra{\Psi}})$
for a bipartition of $\mathcal{H}$ into subsystems $A$ and $B$.

To analyze the power method, we choose $(a, b) = (-1, \Lambda)$, where $\Lambda$ is taken larger than the spectral radius of $\hat{H}$.
Details of how to practically determine $\Lambda$ are discussed in Appendix~\ref{appendix:shifted_scaled_Hamiltonian}.
Based on this setting, taking the $K$th power of $\hat G$ yields
\begin{equation}\label{eq:G^K}
  \hat G^{K}=\sum_{\bm z}\bigl(\Lambda-C(\bm z)\bigr)^{K}\,\ket{\bm z}\bra{\bm z}~.
\end{equation}
Since $\Lambda-C(\bm z)\ge 0$ for all configurations $\bm z$, raising it to the $K$th power exponentially amplifies the relative differences of the eigenvalues between configurations.
In the limit of large $K$, the sum becomes exponentially dominated by the configurations $\{\bm z^\ast\} := \arg \max_{\bm{z}} \bigl(\Lambda - C(\bm{z})\bigr)$,
\begin{equation}\label{eq:lim_coef_G^K}
\lim_{K\to\infty}
\frac{\bigl(\Lambda-C(\bm z)\bigr)^{K}}{\sum_{\bm z'}\bigl(\Lambda-C(\bm z')\bigr)^{K}}
=
\begin{cases}
1/\gamma & \text{if } \bm z\in\{\bm z^\ast\},\\[2pt]
0 & \text{otherwise},
\end{cases}
\end{equation}
where $\gamma=|\{\bm z^\ast\}|$ is the degeneracy (number) of the ground-state configurations of $\hat H$.
From Eqs.~\eqref{eq:G^K} and \eqref{eq:lim_coef_G^K}, $\hat G^{K}$ converges to a sum of $\gamma$ rank-one projectors $ \{ \ket{\bm z^\ast}\bra{\bm z^\ast} \}$ in the limit $K \to \infty$ up to an overall scalar factor, implying it has a bond dimension of, \textit{at most}, $\gamma$.
Specifically, the local projectors for a $d_i$-ary string $\bm z^\ast = z_1^\ast z_2^\ast \cdots z_N^\ast$ are given by
\begin{equation}
  \ket{\bm z_i^\ast}\!\bra{\bm z_i^\ast}
  = \frac{1}{d_i}\sum_{k=0}^{d_i-1} t_{i,k} \hat Z^{(k)}_{i}~,
\end{equation}
with coefficients $t_{i,k} = \bra{\bm z_i^\ast}\hat{Z}^{(k)}_{i}\ket{\bm z_i^\ast}$.

To estimate how large $K$ is required to achieve a target probability $p$ of obtaining the optimal solution from the final state $\ket{\Psi}$, we consider a simplified cost function with the following structure:
\begin{equation}
C({\bm z}) = \left\{
\begin{matrix}
\lambda_{0} & ({\bm z} \in \{ \bm z^\ast\}) \\
\lambda_0 + \Delta & {\rm otherwise}
\end{matrix}
\right.,
\end{equation}
where $\Delta > 0$ is a small constant.
This setup corresponds to a worst-case scenario in which the spectral gap is minimal and the number of suboptimal configurations is maximal, thereby providing an upper bound on the required value of $K$.
In this setting, we first introduce the suppression factor
\begin{equation}\label{eq:eta}
    \eta(\Lambda) = \frac{\Lambda- (\lambda_0 + \Delta)}{\Lambda-\lambda_0}~,
\end{equation}
which quantifies the relative amplitude between the ground state with eigenvalue $\lambda_0$ and a suboptimal configuration at $\lambda_0+\Delta$.
Then, the probability $p$ of sampling the optimal configurations from the state $\ket{\Psi}$ is
\begin{equation}
    p=\frac{\gamma}{\gamma+\left(\mathcal{D}-\gamma \right) \eta(\Lambda)^{2 K}}~,
\end{equation}
where $\mathcal{D}$ is the total dimension of $\mathcal{H}$.
Rearranging the equation gives
\begin{equation}\label{eq:K_eta}
K=\frac{\log \left(\frac{p}{1-p} \right)+\log \left(\frac{\mathcal{D}-\gamma}{\gamma} \right)}{2 \log \left(\frac{1}{\eta(\Lambda)}\right)}~.
\end{equation}
This expression provides a quantitative estimate for the multiplier $K$ needed to amplify the amplitude of optimal configurations such that it appears with probability at least $p$.

We analyze the asymptotic scaling behavior of the required $K$ under the conditions $\mathcal{D} \gg \gamma$, 
$\Delta \ll \Lambda - \lambda_0$, and $p \to 1$.
\begin{equation}\label{eq:asymptotic_scaling}
K = O\left( \frac{\Lambda - \lambda_0}{\Delta} \cdot \log\left( \frac{\mathcal{D}}{1 - p} \right) \right)~.
\end{equation}
This result shows that $K$ scales inversely with the spectral gap $\Delta$, and it also grows logarithmically with both the Hilbert space dimension $\mathcal{D}$, 
which corresponds to the number of possible configurations, 
and the inverse of the failure probability $(1-p)$.
Furthermore, it scales proportionally to the spectral range $(\Lambda - \lambda_0)$, indicating that a tight choice of $\Lambda$ is always favorable.
Hence, for problems with small energy gaps or high target success probability, correspondingly larger values of $K$ are required to effectively amplify the optimal contribution.

\section{Method}\label{sec:method}

We have discussed the quantum-inspired approach for solving COPs in Sec.~\ref{subsec:theoretical_analysis}. In practice, the feasibility of this approach hinges on the computational ability to efficiently construct and numerically handle the powered operator $\hat{G}^K$. To achieve this end, we make use of a range of well-established tensor network methods.

First, we construct an MPO for $\hat G$:
\begin{equation}
    \hat{G} = 
    \sum_{\{ \bm{\alpha} \}} 
    \left(\hat{\mathbf{W}}_{1}\right)_{\alpha_0, \alpha_1} 
    \left(\hat{\mathbf{W}}_{2}\right)_{\alpha_1, \alpha_2} 
    \cdots 
    \left(\hat{\mathbf{W}}_{N}\right)_{\alpha_{N-1}, \alpha_N}~,
\end{equation}
where each $\hat{\mathbf{W}}_{i}$ is a local MPO tensor acting on site $i$ 
and carrying virtual indices $(\alpha_{i-1}, \alpha_i)$, with boundary conditions $\alpha_0 = \alpha_N = 1$.
We describe this MPO construction in Appendix~\ref{appendix:mpo_construction}.

Hereafter, $R$ denotes the bond dimension of the resulting MPO of $\hat{G}$, while $\chi$ is used as a numerical truncation parameter that sets the maximum bond dimension in the computation of $\hat{G}^K$.
In the worst case, $R$ is bounded by the
number of operator terms $N_P$ and the operator-Schmidt rank of
$\hat{G}$ across all bipartitions of the system.

In practice, however, $R$ is often substantially suppressed due to the structure of the interactions.
For example, in general QUBO instances with $N$ variables, 
the maximum bond dimension of $\hat{G}$ is strictly bounded as $R \leq \min(\lfloor N/2 \rfloor + 2, r + 2)$ for completely arbitrary graphs where $r$ is the rank (number of non-zero singular values) of the $ N\times N$ coupling matrix.
Further details of the bound on $R$ are discussed in Appendix~\ref{appendix:mpo_construction}.

Once the MPO representation of $\hat{G}$ is obtained, we perform power iteration using the zip-up algorithm~\cite{Stoudenmire_2010}, a practical scheme for MPO–MPO and MPO–MPS multiplications.
One may also consider the fitting algorithm as an alternative.
It should be noted, however, that while both share the same asymptotic order of complexity, the fitting algorithm is a variational method that is sensitive to the initial guess and often necessitates multiple sweeps.

Throughout the multiplication procedures, the bond dimensions are controlled by performing a full SVD at each step.
We retain at most $\chi$ singular values and additionally discard those smaller than a relative threshold of $\varepsilon$.
In this paper, a fixed threshold of $\varepsilon = 10^{-15}$ is maintained across all calculations, including the DMRG results presented later.
The resulting powered-MPO is always renormalized after each multiplication to prevent numerical overflow or underflow.
Efficient normalization in MPO is achieved by rescaling the singular-value tensor $\bm{\sigma}$ in the mixed canonical representation of the MPO, as shown in Appendix~\ref{appendix:mpo_construction} [Eq.~\eqref{eq:mixed_canonical_form}].

In the standard power-iteration approach, the operator $\hat{G}$ is applied $K$ successive times to the resulting MPO.
The computational cost in this case scales as $O(N d^2 R^2 \chi^2 + N d^2 R \chi^3 (K - 1))$. 
To significantly accelerate the construction of $\hat{G}^K$, we can instead repeatedly multiply the resulting MPO with itself.
This allows us to reach $K = 2^m$ with only $m$ multiplications, although with a larger scaling of $\chi$ of $O(N d^2 R^2 \chi^2 + N d^2 \chi^4 (m - 1))$. 
From here on, we refer to these two contraction sequences as a ``linear" and ``doubling" schedule respectively.

The resulting MPO of $\hat{G}^K$ can be embedded into an MPS via Eq.~\eqref{eq:Psi}.
Since the state $\ket{+}^{\otimes N}$ admits an MPS representation with bond dimension one,
this embedding can be done in $O(Nd^{2}\chi^{2})$ time.
Finally, we perform perfect sampling on the MPS $\ket{\Psi}$ with a computational cost of $\mathcal{O}(N d \chi^3)$~\cite{PhysRevB.85.165146}, per sample, 
which yields a set of $\bm{z}$ whose amplitudes are biased toward low-energy states of $\hat{H}$, 
as described in Sec.~\ref{subsec:theoretical_analysis}.

\section{Numerical Benchmarks and Performance Evaluation}\label{sec:benchmark}

\begin{figure*}[ht]
  \centering
  \includegraphics[clip,width=1.0\textwidth]{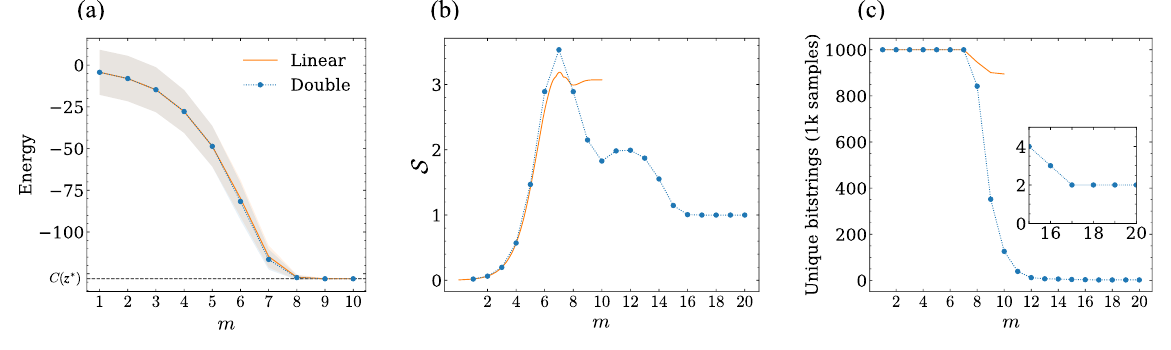}
    \caption{
    Panel (a) shows the mean energy and its standard deviation obtained from 1,000 samples of $\ket{\Psi}$ for a specific instance with $(D,L)=(2,10)$.
    The Label “Linear” denotes a linear schedule in which the power $K$ is increased step by step by repeatedly multiplying by $\hat{G}$, whereas label “Double” means using a doubling schedule $K=2^m$ by multiplying the result of the MPO-MPO contraction with itself $m$ times.
    On the horizontal axis, $C(\bm{z}^\ast)$ denotes the optimal cost value of the instance.
    Panel (b) shows the $K~(=2^m)$ dependence of the maximum bipartite entanglement entropy $\mathcal{S}$.
    Panel (c) plots the number of unique bitstrings among $1000$ samples.
    We used $\chi=64$ to generate data for these plots. 
}\label{fig:demonstration}
\end{figure*}

To quantitatively evaluate the performance of our proposed method,
we first conduct numerical experiments on Ising Hamiltonians defined on
$D$-dimensional hypercubic lattices of linear size $L$, 
so that the number of binary variables is $N=L^D$.  
Let us label each site by a coordinate vector
\begin{equation}\label{eq:ising_spin_glass}
    \bm r = (r_1,r_2,\ldots,r_D), \qquad r_{\nu} \in \{0,1,\ldots,L-1\}~.
\end{equation}
The Hamiltonian is given by
\begin{equation}\label{eq:benchmark_hamiltonian}
\hat{H} = \sum_{(\bm r,\bm r^{\prime})\in E} 
J_{\bm r, \bm r^{\prime}}\, \hat{Z}_{\bm r} \hat{Z}_{\bm r^{\prime}}~,
\end{equation}
where $J_{\bm r,\bm r^{\prime}}\in\{-1,+1\}$ are independently assigned random, bimodal variables, and the summation is taken over the set of edges $E$ of the lattice.
In this study, the ground-state energy $C(\boldsymbol{z}^\ast)$ of Eq.~\eqref{eq:benchmark_hamiltonian} is obtained using \texttt{CPLEX}~\cite{cplex}, a high-performance solver for finding exact solutions to mixed-integer programming problems.

In this Hamiltonian, the bond dimension of the MPO is limited by the locality of the pairwise interactions.
We ensure that $R$ grows as $L^{D-1} + 2$ under the mapping of the $D$-dimensional lattice onto a one-dimensional chain:
\begin{equation}
n(\boldsymbol{r}) = 1 + \sum_{\nu=1}^D r_\nu L^{\nu-1}~,
\end{equation}
where $n(\boldsymbol{r}) \in {1, 2, \ldots, N}$ denotes the one-dimensional site index.
This rank is consistent with the operator Schmidt rank argument shown in Appendix~\ref{appendix:mpo_construction}, as the number of interaction terms crossing a bipartition of the chain is proportional to the boundary of the original $D$-dimensional hypercube.
As our goal is to obtain the ground-state energy of Eq.~\eqref{eq:benchmark_hamiltonian}, we adopt $(a,b)=(-1,\Lambda)$ introduced in Appendix~\ref{appendix:shifted_scaled_Hamiltonian} [Eq.~\eqref{eq:extreme_setting}] for the transformation of $\hat{H}$ into $\hat{G}$.

\subsection{Energy Distributions and Entanglement Entropy}\label{subsec:minimum_benchmark}

We begin by presenting the mean energy and its standard deviation of sampled energies for one instance of the lattice with $(D, L)=(2,10)$ in Fig.~\ref{fig:demonstration}(a), evaluated at a fixed bond dimension of $\chi=64$.
As expected, the convergence behavior is similar for both schedules, following the asymptotic scaling in Eq.~\eqref{eq:asymptotic_scaling}.
Specifically, Fig.~\ref{fig:demonstration}(a) crealy demonstrates the resulting MPS reaches a superposition of configurations drawn from a subset of $\{\bm{z}^\ast\}$ by $m=10$.

Furthermore, we investigate the EE (Eq.~\eqref{eq:EE_def}) in the resulting MPS.
For a given bipartition, the EE of $\ket{\Psi}$ can be computed from the singular-values tensor $\bm \sigma$ in the mixed canonical form of the MPO [Eq.~\eqref{eq:mixed_canonical_form}] as follows: 
\begin{equation}
\mathcal{S} = - \sum_i \sigma_i^{2}\,\log_{2}\!\big(\sigma_i^{2}\big)~.
\end{equation}
In Fig.~\ref{fig:demonstration}(b), we report the $K$-dependence of the \emph{maximum} EE over all bipartitions in the MPS.
The linear and doubling schedules yield virtually identical peak structures.
This peaks remain noticeably below the theoretical maximum $\log_{2}\chi=6$ for $\chi=64$, corroborating that the power method with an MPS is feasible with a reasonable bond dimension.

Under the doubling schedule, the EE shows a more pronounced decay after reaching its peak, suggesting that the squared-power iteration can be more sensitive to truncation errors than the linear schedule.
To analyze this entanglement behavior more precisely,
we plot the number of sampled unique bitstrings in Fig.~\ref{fig:demonstration}(c).
This result indicates that at $K=1,024$ ($m=10$), the linear schedule identifies more than $850$ unique ground state configurations from $\{\bm{z}^\ast\}$, whereas the doubling schedule identifies fewer than $200$ within the same number of samples.
Additionally, this value further decreases and fully converges to $2$ by $m=20$ in the doubling schedule, resulting in $\mathcal{S}=1$.
We observed that these two configurations are related by a spin flip, which is expected for even $L$ owing to the global $\mathbb{Z}_2$ spin-flip symmetry.

\subsection{Performance Benchmarking with DMRG}

\begin{figure*}[ht]
  \centering
  \includegraphics[clip,width=1.0\textwidth]{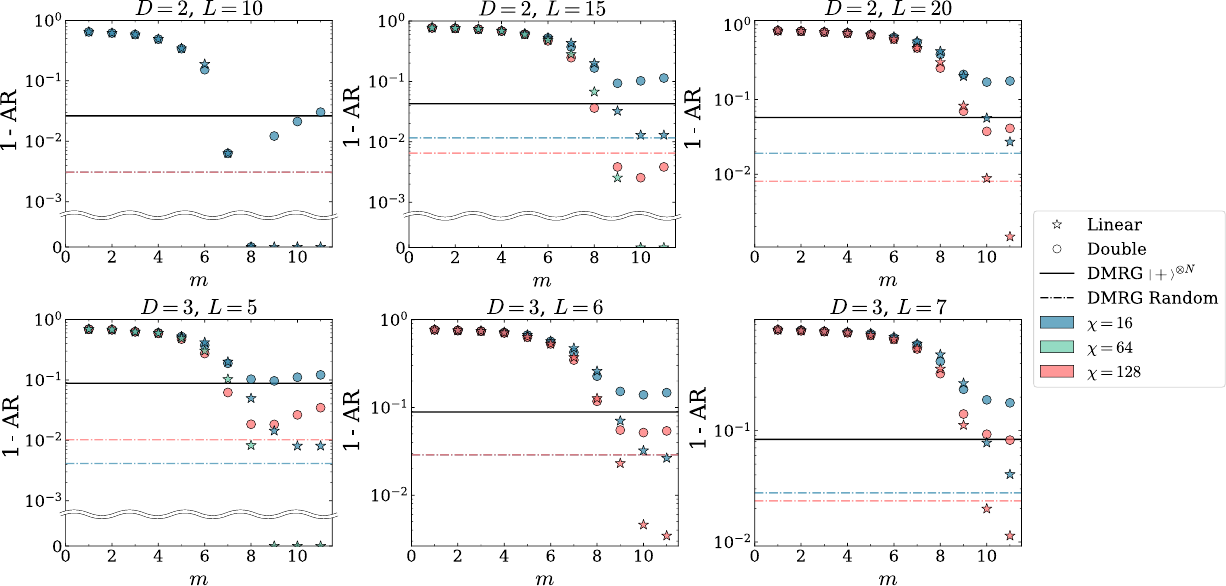}
  \caption{Error, $1 - {\rm AR}$, where ${\rm AR}$ is the mean approximation ratio [Eq.~\eqref{eq:ar}] averaged over five problem instances for powered-MPO method and DMRG; $D$ denotes the spatial dimension and $L$ signifies the linear size of $D$-dimensional hypercubic lattice. 
  Stars and circles represent the powered-MPO results using linear and doubling schedules, respectively.
  Colors indicate the bond dimension used in the calculations.
  The black line represents the DMRG results starting from the initial state $\ket{+}^{\otimes N}$, which remain invariant across the investigated bond dimensions $\chi \in \{16, 64, 128\}$.
  Colored dash-dotted lines indicate the DMRG results starting from random MPS.
  In certain cases the powered-MPO approaches successfully reache the exact solution to the COP  and this is indicated by the broken axis and the data lying exactly on the $x$-axis.}
  \label{fig:main_result}
\end{figure*}

\begin{table*}
\label{tab:ar_results}
  \vspace{2mm}
  \centering
    \caption{
  Mean and standard deviation $(\mu, \sigma)$ of the maximum AR over five instances for all tested lattice dimensions $(D,L)$.
  For the powered-MPO, we report the maximum AR attained at the peak step $m$ (identified within $m \in [1, 11]$) and then average these peak AR values over the five instances.
  For DMRG initialized with the $\ket{+}^{\otimes N}$ state, no specific numerical bond dimension is assigned (denoted simply as $\chi$), as the results remain invariant across $\chi \in \{16, 64, 128\}$.
  In the case of using random initial MPS for DMRG, ten independent states were prepared with an initial bond dimension matching the target $\chi$ for calculations.
  These states were generated using ten fixed random seeds shared across run.
  }
  
  \label{tab:ar_results}
  \resizebox{\textwidth}{!}{%
  \begin{tabular}{|c|c|c|c|c|c|c|c|c|}
    \Xhline{1.1pt}
    \multirow{2}{*}{\textbf{State}} & \multirow{2}{*}{\textbf{Method}} & \multirow{2}{*}{$\chi$}
      & \multicolumn{3}{c|}{$D=2$} & \multicolumn{3}{c|}{$D=3$} \\ \cline{4-9}
    & & & $L=10$ & $L=15$ & $L=20$ & $L=5$ & $L=6$ & $L=7$ \\
    \Xhline{0.9pt}
    \multirow{7}{*}{$\ket{+}^{\otimes N}$}
    & \multirow{3}{*}{Linear} & 16  & (1.000, 0.000) & (0.987, 0.005) & (0.973, 0.010) & (0.992, 0.008) & (0.975, 0.005) &  (0.960, 0.009) \\ \cline{3-9}
    &                              & 64  & - & (1.0000, 0.0000) & (0.984, 0.004) & (1.000, 0.000) & (0.989, 0.006) & (0.984, 0.007) \\ \cline{3-9}
    &                              & 128 & - & - & (0.999, 0.002) & - & (0.997, 0.005) 
 & (0.989, 0.004) \\ \cline{2-9}
    
    & \multirow{3}{*}{Double} & 16  & (1.000, 0.000) & (0.908, 0.013) & (0.829, 0.014) & (0.906, 0.027) & (0.865, 0.020) & (0.822, 0.008) \\ \cline{3-9}
    &                              & 64  & - & (0.987, 0.005) & (0.954, 0.007) & (0.971, 0.021) & (0.936, 0.018) & (0.870, 0.008) \\ \cline{3-9}
    &                              & 128 & - & (0.994, 0.009) & (0.963, 0.007) & (0.984, 0.016) & (0.949, 0.015)
 & (0.918, 0.010) \\ \cline{2-9}
    & DMRG                         & $\chi$  & (0.974, 0.026) & (0.957, 0.024) & (0.942, 0.015) & (0.912, 0.024) & (0.911, 0.023) & (0.917, 0.011) \\ \hline
    \multirow{3}{*}{Random}
    & \multirow{3}{*}{DMRG}        & 16  & (0.997, 0.007) & (0.988, 0.010) & (0.981, 0.006) & (0.996, 0.006) & (0.971, 0.006) &  (0.972, 0.005) \\ \cline{3-9}
    &                              & 64  & (0.997, 0.007) & (0.991, 0.014) & (0.990, 0.006) & (0.996, 0.006) & (0.975, 0.006) & (0.975, 0.003) \\ \cline{3-9}
    &                              & 128 & (0.997, 0.007) & (0.994, 0.011) & (0.992, 0.005) & (0.990, 0.013) & (0.971, 0.013) & (0.977, 0.006) \\
    \Xhline{1.1pt}
  \end{tabular}
  }
\end{table*}

To further evaluate the powered-MPO method, we benchmark its performance against DMRG \cite{morais2025comparative}, the gold standard for high-precision MPS optimization.
Here, we extend the evolution to $K=2^{11}$ steps ($m=11$) across various bond dimensions $\chi \in \{16, 64, 128\}$.
For a rigorous comparison, these same bond dimension limits are applied to the two-site DMRG algorithm as the maximum bond dimensions.
Furthermore, to account for the sensitivity of DMRG to initial MPS to be input, we perform calculations using both the uniform superposition $\ket{+}^{\otimes N}$ and ten independent random initial MPS.
Each tensor in the random MPS is initialized as a random isometry with a bond dimension matching the maximum $\chi$ allowed in the calculations.
We perform DMRG sweeps until the relative energy change drops below a threshold of $10^{-13}$.

All algorithms are tested on hypercubic lattices with dimensions $(D,L) \in \{ (2,10),(2,15),(2,20),(3,5),(3,6),(3,7) \}$, using $5$ random instances for each.
We benchmark performance by measuring the approximation ratio (AR), defined as
\begin{equation}\label{eq:ar}
  \mathrm{AR} := \frac{\min_{\bm{z}\in\mathcal{Z}} C(\bm{z})}  {C(\bm{z}^\ast)}~,
\end{equation}
where $\mathcal{Z}$ is a set of 1000 sampled bitstrings from the resulting MPS. 
We summarize the results in Figs.~\ref{fig:main_result} and Table~\ref{tab:ar_results}, plotting the ``error" $1-{\rm AR}$ on a log-scale in the former for visual clarity.

We first compare the linear and doubling schedules within the powered-MPO framework, as shown in Fig.~\ref{fig:main_result}. 
At the earliest stages, both schedules produce comparable AR values.
During the intermediate stages, the doubling schedule shows a more rapid improvement in AR before eventually saturating or even beginning to decrease slightly.  We extended the doubling calculations to $m = 20$ and confirmed that they ultimately converge to a lower AR than from its peak value around $8 \leq m \leq 11$.
In contrast, the linear schedule sustains gradual yet consistent improvements beyond the saturation point reached by the doubling schedule, resulting in higher final accuracy which appears systematically improveable with both $m$ and $\chi$ --- with the optimal solution being reached in the smaller system sizes.
It appears that prolonged linear evolution may eventually reach the optimal AR in the larger cases, even with limited bond dimensions.

These results reveal the complementary strengths of the two schedules.
The doubling schedule provides rapid initial stages, whereas the linear schedule enables continued, more stable improvements at later stages.
This fact naturally motivates a hybrid strategy that transitions from fast initial filtering to high-precision refinement.
Notably, the computational cost per iteration also differs significantly.
The doubling schedule scales as $O(\chi^4)$ (assuming $\chi \geq R$), while the linear schedule scales as $O(R\chi^3)$ and thus matches the computational scaling of DMRG.

In our DMRG calculations, shown in Fig. \ref{fig:main_result}, we find DMRG tends to become trapped in local minima, especially when started from the uniform state $\ket{+}^{\otimes N}$.
Specifically, we find it converges to the same solution regardless of the bond dimension; increasing $\chi$ from $\{ 16, 64, 128 \}$ yields no improvement. 
In contrast, the powered-MPO method shows a systematic improvement in AR as $\chi$ is increased. 
These results suggest that our method more effectively explores the relevant Hilbert space from the starting point of $\ket{+}^{\otimes N}$, whereas the sweeping optimization of DMRG is more susceptible to local trapping.

On the other hand, we do find that random initial states helps DMRG circumvent some of these local traps, and improve on the AR values it obtains.
As shown in Table~\ref{tab:ar_results}, however, the improvement in AR for random-start DMRG still remains relatively marginal as $\chi$ is increased from $16$ to $128$, suggesting that the variational sweeping is limited by the cost landscape encountered in its initial stages.
For several instances where DMRG remains stagnant in near-optimal configurations, the powered-MPO method successfully provides the maximum AR once a sufficient bond dimension is used.

Despite this, DMRG can remain a compelling option in specific contexts, particularly when seeking high-quality solutions with limited walltime and resources.
While the computational cost per sweep in DMRG is identical to power iteration using a linear schedule, the DMRG achieves convergence in fewer sweeps than the number of MPO-MPO contractions in the powered MPO method.

\subsection{Benchmark for higher-order Ising spin glass on heavy-hexagonal architecture}

\begin{figure*}[t]
  \centering
  \includegraphics[clip,width=1.0\textwidth]{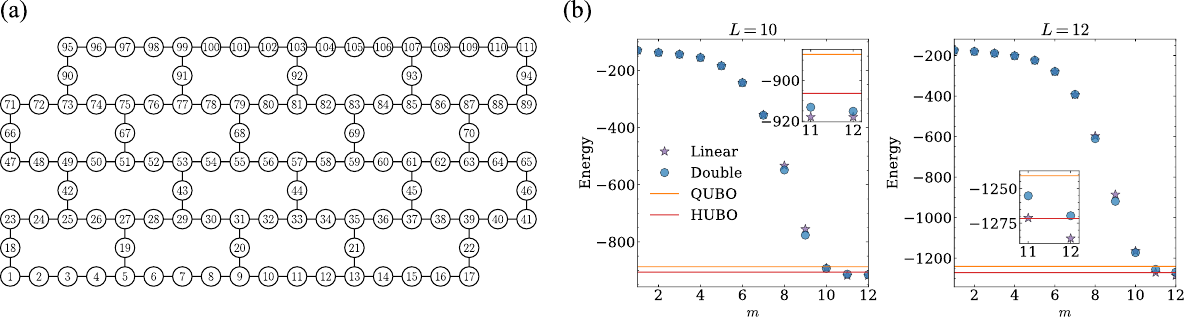}
  \caption{(a) Schematic of a heavy-hexagonal lattice with size $L=4$, consisting of $L \times L$ unit cells.
  The vertex indices denote the one-dimensional site ordering adopted for the MPO representation of the Hamiltonian. 
  (b) Average minimum energy obtained with the powered-MPO method as a function of $m$, compared with SA results.
  Markers represent mean values across five problem instances; error bars (standard deviation) are omitted for clarity.
  The insets show a magnified view of the results for $m > 10$.
  All MPO calculations were performed with a maximum bond dimension of $\chi=64$.}
  \label{fig:heavyhex_results}
\end{figure*}

With the recent advances in quantum computing, QAOA has lately been benchmarked to evaluate its practical usability on current quantum hardware.
Among many demonstrations, Ref.~\cite{QA_vs_QAOA_127} designed HUBO problems including up to cubic interaction terms tailored to the IBM heavy-hex architecture, and compared their performance with QA on D-Wave quantum annealers.

While QA requires order reduction by introducing auxiliary variables, quantum computers can natively implement higher-order interactions using sequences of CNOT gates.
The rugged energy landscapes of HUBO exacerbate the risk of trapping classical heuristics in local minima.
Consequently, HUBO is regarded as a problem class that can particularly benefit from the use of quantum computers.
In fact, numerical studies demonstrate that QAOA can extract optimal solutions for the problem in Ref.~\cite{QA_vs_QAOA_127} by employing deeper QAOA schedules with parameters trained on, and transferred from, smaller problem instances~\cite{pelofske2025evaluatinglimitsqaoaparameter}.

We apply the powered-MPO to the HUBO problems in Ref.~\cite{QA_vs_QAOA_127}, formulated as 
\begin{equation}\label{eq:heavyhex_HUBO}
\begin{split}
    \hat{H} &= \sum_{v \in V} d_v \hat{Z}_v
    + \sum_{(i, j) \in E} d_{i, j}\,\hat{Z}_i \hat{Z}_j \\
    &+ \sum_{\ell \in W} d_{\ell, n_1(\ell), n_2(\ell)}\,\hat{Z}_\ell \hat{Z}_{n_1(\ell)} \hat{Z}_{n_2(\ell)}~,
\end{split}
\end{equation}
where $V = \{0, \dots, N-1\}$ and $E \subset V \times V$ denote the sets of vertices and edges, respectively, and $W$ is the subset of vertices with degree two.
We set the coeffcients $d_v$, $d_{i,j}$, and $d_{\ell, n_1(\ell), n_2(\ell)}$ randomly from $\{ \pm1\}$. 
Let $L$ be the number of unit cells along one side of the heavy-hexagonal lattice.
The lattice is obtained from a honeycomb lattice consisting of $L \times L$ cells by inserting one additional vertex at the midpoint of every edge; therefore, $|V| = 5L^2 + 8L - 1$ and $|E| = 6L^2 + 8L - 2$.
Figure~\ref{fig:heavyhex_results} (a) illustrates an example of a heavy-hexagonal lattice with $L=4$.

In the experiments, we compare the performance of the powered-MPO method against SA, one of the most widely used classical heuristics for searching the ground state of Ising Hamiltonians.
We employ \texttt{OpenJij}~\cite{nishimura_2025_17345408} to run SA, since it natively supports HUBO formulations.
Additionally, we also study SA applied to the order-reduced QUBO formulation using \texttt{dimod}~\cite{dimod2025}, 
with a penalty strength of $2.0$, which ensures that the optimal solution of the QUBO model coincides with that of Eq.~\eqref{eq:heavyhex_HUBO}.
In order to construct an MPO of cost Hamiltonians, we arrange the vertices into a one-dimensional chain following the indices shown in Fig.~\ref{fig:heavyhex_results} (a).
We observe that the maximum MPO bond dimension for $\hat{G}$ is effectively bounded by $L+6$ due to the sparsity of the lattice, 
even though 
$|W|$ in Eq.~\eqref{eq:heavyhex_HUBO} also scales as $L^2$, like $|V|$ and $|E|$.

Fig.~\ref{fig:heavyhex_results}(b) presents the average minimum energies across five instances for $L=10$ and $L=12$ heavy-hex lattices, where each data point is derived from 1,000 samples per method.
For the powered-MPO method, we employed a maximum bond dimension of $\chi=64$.
At $L=10$, powered-MPO method with $\chi=64$ already outperforms SA in both its QUBO and HUBO formulations.
In the $L=12$ case, while the doubling and linear schedules exhibit nearly identical performance up to $m=7$, the linear evolution eventually proves more robust, becoming the only method to surpass the HUBO solutions upon reaching $m=12$.
These results suggest that the powered-MPO method remains robust against rugged energy landscapes, consistently yielding accurate solutions as long as the bond dimension $\chi$ is adequate.

We emphasize, however, that the performance of SA can often be improved by careful tuning of hyperparameters or by adopting algorithmic variants tailored to specific problem classes.
Similarly, the powered-MPO framework can be augmented by incorporating heuristic strategies.
For instance, applying $\prod_{i=1}^{N} \exp(-\beta \hat{X}_i)$ to identified classical configurations effectively warm-starts the optimization.
This operation generates a superposition whose amplitudes decay exponentially with the Hamming distance from the input, where $\beta \in [0, \infty)$ controls the degree of localization.
Crucially, this initialization introduces no additional entanglement as it consists only of single-qubit rotations.

\section{Conclusion and Outlook}\label{sec:conclusion}

In this work, we propose a strategy for solving combinatorial optimization problems based on power iteration of the cost Hamiltonian within the MPO framework, followed by sampling from the resulting quantum state.
This pipeline drastically amplifies the weight of ground-state configurations through tensor contractions and SVD-based truncation, so that the sampling distribution becomes sharply peaked around optimal bitstrings as the exponent on the cost function increases.
We compared our approach to DMRG and SA, both of which often become trapped in local minima leading to lower quality solutions than those found within our framework.
When a linear schedule is used, our method has identical scaling in bond dimension to the DMRG algorithm and shows more systematic improvement when increasing it - although it requires a lot more steps than DMRG does to converge.
To reach exact solutions with our method one must still overcome high entanglement barriers, and the scaling of the required bond dimension to achieve this across various types of problems requires systematic clarification in future work.
In particular, implementing MPI-level parallelism would enable bond dimensions reaching $\chi \sim \mathcal{O}(10^3)$~\cite{PhysRevB.101.235123,PhysRevB.110.085149}, and we also observed that GPUs (which were used to obtain some of the data in this work) can significantly accelerate the tensor contractions ~\cite{10.1145/3696465,Lyakh2022ExaTN,Vallero_2026, rudolph2025simulatingsamplingquantumcircuits}, which form the computational workhorse of our method.

As a future approach to keeping the bond dimension tractable on large instances of more complex lattice topologies, we could employ lattice-specific TN architectures.
In particular, extending our method to tree TNs is straightforward, as they contain no loops~\cite{Szalay_2015,milbradt2026efficientapplicationtensornetwork}.
Moreover, we can adopt structural search algorithms on TTNs to find a tree structure tailored to a specific instance, thereby reducing the required bond dimensions~\cite{PhysRevResearch.5.013031,watanabe2025ttnopttreetensornetwork}.
Regarding loopy TNs, we tested projected entangled-pair operators (PEPOs) on square and heavy-hexagonal lattices for the Hamiltonian in Eqs.~\eqref{eq:benchmark_hamiltonian} and \eqref{eq:heavyhex_HUBO} in Sec.~\ref{sec:benchmark}, and confirmed that the bond dimension of the associated PEPOs remained constant.
Direct contraction between two PEPOs is required at intermediate stages, however, and we observed that the resulting PEPO was fragile under truncation based on belief propagation techniques~\cite{Tindall_2023}, suggesting the need for more accurate truncation methods~\cite{PhysRevB.98.085155}.
Nonetheless, once the amplified projected entangled-pair state (PEPS) is obtained, we can sample from this state accurately by relying on recent developments for sampling from PEPS in the TN community~\cite{PhysRevB.85.165146, rudolph2025simulatingsamplingquantumcircuits}.
Developing more stable and efficient PEPO/PEPS contraction techniques are therefore an important direction for future work.

The significant quality of the results obtained by our powered-MPO approach also means they could serve as a tensor network-based baseline when attempting to use a quantum computer (QC) to demonstrate quantum utility / advantage for solving COPs. Moreover, we could imagine attempting to identify the shallowest, measurement-based circuit which implements the action of our non-unitary cost-function and can then be implemented on a quantum computer to solve the given COP.
Current limitations on qubit counts in trapped ion QCs and accurately implementing non-local operations in superconducting-based architectures, however, represent obstacles to this approach.

\section*{Acknowledgments}
We acknowledge the use of \texttt{ITensor} ecosystem for the Tensor Network simulations~\cite{ITensor} in this work.
This work is partially supported by KAKENHI Grant Numbers JP21H04446, JP21H05182, JP21H05191, JP24K16978, and 25KJ1773 from JSPS of Japan and by the New Energy and Industrial Technology Development Organization (NEDO) (Grant No. JPNP20017).
We also acknowledge support from MEXT Q-LEAP Grant No. JPMXS0120319794, and from JST COI-NEXT No. JPMJPF2014, ASPIRE No. JPMJAP2319, and CREST No.JPMJCR24I1 and JPMJCR24I3, and the COE research grant in computational science from Hyogo Prefecture and Kobe City through Foundation for Computational Science. JT is grateful for ongoing support through the Flatiron Institute, a division of the Simons Foundation. 

\bibliographystyle{quantum}
\bibliography{main}

\appendix
\section{Shifted and scaled cost Hamiltonian}\label{appendix:shifted_scaled_Hamiltonian}

Let $\{\lambda_j\}_{j=0}^{\mathcal{D}-1}$ denote the eigenvalues of $\hat H$ arranged in ascending order, where $\mathcal{D} = \prod_i d_i$ is the total dimension of the Hilbert space $\mathcal{H}$.
We can introduce simple examples of such transformations:
\begin{equation}\label{eq:simple_G}
\hat G =
\begin{cases}
\hat H - \lambda_{0}~\mathbb I & \text{maximize } C(\bm z) \\
\lambda_{\mathcal{D}-1}~\mathbb I - \hat H & \text{minimize } C(\bm z)
\end{cases}~.
\end{equation}

In practice, however, the target eigenvalues of $\hat{H}$ or equivalently the target value of $C(\bm{z})$ may be unknown.
Therefore, we introduce a shift parameter $\Lambda$ such that
\begin{equation}\label{eq:conditions_on_Lambda}
\lambda_{0} \ge -\Lambda, \qquad \lambda_{\mathcal{D}-1} \le \Lambda~.
\end{equation}
With this definition, we can always adopt
\begin{equation}\label{eq:extreme_setting}
(a,b) =
\begin{cases}
(1,\ -\Lambda) & \text{maximize } C(\bm z)~, \\
(-1,\ \Lambda) & \text{minimize } C(\bm z)~.
\end{cases}
\end{equation}
To satisfy the condition in Eq.~\eqref{eq:conditions_on_Lambda}, a conservative but always valid choice is
\begin{equation}
\Lambda = \sum_{j=1}^{N_P} |h_j|\,\|\hat O_j\|~,
\end{equation}
where $\| \cdot \|$ denotes the operator norm.

\section{Procedure for constructing the MPO representation}\label{appendix:mpo_construction} 

The transformed Hamiltonian $\hat{G}$ in Eq.~\eqref{eq:G} can be written as 
\begin{equation}
    \hat{G} = \sum_{\{\bm{p}\}} g_{p_1\cdots p_{N}} \hat{Z}^{(p_1)}_1 \cdots \hat{Z}^{(p_{N})}_{N}
    \label{eq:operator_G}
\end{equation} 
with the operator basis $\mathcal{O}$ in Eq.~\eqref{eq:operator_basis}, where $g_{p_1\cdots p_N} \in \mathbb{R}$ and each $\hat{Z}^{(p_i)}_i$ is chosen from $\mathcal{P}_i$ defined in Eq.~\eqref{eq:local_operator_basis}, with $p_i \in \{ 0, \ldots, d_i - 1 \}$.
Recall that $\hat{G}$ is positive semidefinite.
We denote by $N_{P_0}$ the number of nonzero coefficients $g_{p_1 \cdots p_{N}}$.
This representation allows us to construct the corresponding MPO by sequentially applying SVD from one end of the chain.

Let us now define an index subset $P'_{1} \subset \{p_2,\ldots,p_{N}\}$ with $|P'_{1}| \leq N_{P_0}$ such that the tensor product $\{p_1\} \times P'_1$ captures all nonzero coefficients $g_{p_1 \cdots p_{N}}$.
Such a subset $P'_{1}$ can be determined numerically in a straightforward manner.
We can then define a $d_1 \times N_{P_1}$ matrix with $N_{P_1} \leq N_{P_0}$,
\begin{equation}
    \mathsf{M}^{(1)} = \left\{ \mathsf{m}^{(1)}_{p_1, p_1'} = g_{p_1, p_1'} \right\}_{0 \leq p_1 \leq d_1-1,\ p_1' \in P'_1}~.
\end{equation}
Applying SVD to $\mathsf{M}^{(1)}$, we obtain 
\begin{equation}
    \mathsf{m}^{(1)}_{p_1, p_1'} = \sum_{\alpha_1} u_{p_1,\alpha_1}^{(1)} \lambda_{\alpha_1}^{(1)} v_{\alpha_1,p_1'}^{(1)}~, 
\end{equation} 
where the singular values $\lambda_{\alpha_1}^{(1)}$ are sorted in descending order and may be truncated if they are numerically negligible. 
In terms of the MPO representation, the first tensor is given by
\begin{equation} 
    \left( \hat{\mathbf{L}}_1 \right)_{\alpha_0,\alpha_1} = \sum_{p_1} u_{p_1,\alpha_1}^{(1)} \hat{Z}^{(p_1)}_1~,
\end{equation} 
with the boundary condition $\alpha_0 = 1$.
We then define $g_{\alpha_1, p_1'}^{(1)} = \lambda_{\alpha_1}^{(1)} v_{\alpha_1,p_1'}^{(1)}$ to construct the second MPO tensor.

Next, proceeding similarly, we define a minimal subset $P'_2 \subset \{p_3,\ldots,p_{N}\}$ such that the set $\{\alpha_1\} \times \{p_2\} \times P'_2$ spans all nonzero elements of $g_{\alpha_1, p_1'}^{(1)}$.
We then reshape $g_{\alpha_1, p_1'}^{(1)}$ and construct the third-order tensor $\mathsf{m}^{(2)}_{\alpha_1, p_2, p_2'}$ with $p_2' \in P'_2$, and treat it as a matrix with row indices $(\alpha_1, p_2)$ and column index $p_2'$.
Then we apply SVD to obtain
\begin{equation}
    \mathsf{m}^{(2)}_{\alpha_1, p_2, p_2'} = \sum_{\alpha_2} u_{\alpha_1, p_2, \alpha_2}^{(2)} \lambda_{\alpha_2}^{(2)} v_{\alpha_2, p_2'}^{(2)}~,
\end{equation}
with $g_{\alpha_2, p_2'}^{(2)} = \lambda_{\alpha_2}^{(2)} v_{\alpha_2, p_2'}^{(2)}$.
The second MPO tensor is given by
\begin{equation}
    \left( \hat{\mathbf{L}}_2 \right)_{\alpha_1, \alpha_2} = \sum_{p_2} u_{\alpha_1, p_2, \alpha_2}^{(2)} \hat{Z}^{(p_2)}_2~.
\end{equation}
By repeating this procedure recursively, we obtain tensors $\hat{\mathbf{L}}_i$ up to $i = N-1$. The final tensor at site $N$ is constructed as
\begin{equation}
    \left( \hat{\mathbf{W}}_{N} \right)_{\alpha_{N-1}, \alpha_{N}} = \sum_{p_{N}} g_{\alpha_{N-1}, p_{N}}^{(N-1)} \hat{Z}^{(p_{N})}_{N}~,
\end{equation}
with the boundary condition $\alpha_N = 1$.
Then we finally obtain the MPO
\begin{equation}
\begin{split}
    \hat{G} =
    \sum_{\{ \bm{\alpha} \}} 
    \left(\hat{\mathbf{L}}_{1}\right)_{\alpha_0, \alpha_1}
    &\cdots 
    \left(\hat{\mathbf{L}}_{N-1}\right)_{\alpha_{N-2}, \alpha_{N-1}} \\
    &\times
    \left(\hat{\mathbf{W}}_{N}\right)_{\alpha_{N-1}, \alpha_N}~.
\end{split}
\end{equation}
Here, the tensors $\hat{\mathbf L}_i$ satisfy the isometric condition with respect to the right index $\alpha_i$:
\begin{equation}
    \sum_{z_i} \sum_{\alpha_{i-1}}\bra{ z_i}\left(\hat{\mathbf{L}}_i\right)_{\alpha_{i-1}, \alpha_i}^{\dagger}\left(\hat{\mathbf{L}}_i\right)_{\alpha_{i-1}, \alpha^{\prime}_i}\ket{z_i} = \delta_{\alpha_{i}, \alpha^{\prime}_{i}}~.
\end{equation}

After constructing the left-canonical tensors up to site $c$,
the right-canonical $\hat{\mathbf{R}}_{c+1},\ldots,\hat{\mathbf{R}}_{N}$ are obtained in an analogous manner by applying SVDs from the right end of the chain, which yields the singular values $\lambda^{(c)}_{\alpha_c}$ at the orthogonality center.
Using these left- and right-canonical tensors,
the MPO can be written in mixed canonical form as  
\begin{equation}\label{eq:mixed_canonical_form}
\begin{split}
\hat{G} 
&= \sum_{\{ \bm{\alpha} \}} 
    \left(\hat{\mathbf{L}}_{1}\right)_{\alpha_0, \alpha_1} \cdots
    \left(\hat{\mathbf{L}}_{c}\right)_{\alpha_{c-1}, \alpha_c} \\
&\quad \times~
    \lambda^{(c)}_{\alpha_c} \times
    \left(\hat{\mathbf{R}}_{c+1}\right)_{\alpha_c, \alpha_{c+1}}
    \cdots
    \left(\hat{\mathbf{R}}_{N}\right)_{\alpha_{N-1}, \alpha_N}~,
\end{split}
\end{equation}
where each
$\mathbf{R}_i$ satisfies the right-canonical condition
\begin{equation}
    \sum_{z_i} \sum_{\alpha_{i}}
    \bra{z_i}\left(\hat{\mathbf{R}}_i\right)_{\alpha_{i-1}, \alpha_i} \left(\hat{\mathbf{R}}_i\right)_{\alpha_{i-1}^{\prime},\alpha_i}^{\dagger}\ket{z_i} =\delta_{\alpha_{i-1}, \alpha_{i-1}^{\prime}}~.
\end{equation}

Using the SVD-based MPO construction described above, the bond
dimension at each cut is bounded by the operator Schmidt rank of
$\hat{G}$ across that bipartition.
Let us discuss the bound of bond dimension in the QUBO operator
\begin{equation}
    \hat{H} = \sum_{i = 1}^{N}\sum_{\substack{j > i}}^{N}
    M_{i,j}\,\hat{Z}_{i}\hat{Z}_{j}~,
    \label{eq:QUBOMPO}
\end{equation}
where $M_{i,j}\in\mathbb{R}$ are real coupling coefficients.
Since $\hat{G}$ differs from the original cost Hamiltonian $\hat{H}$ only by a global scaling and a shift, the structure relevant to the Schmidt decomposition is entirely carried by $\hat{H}$, while the shift term can contribute only a trivial rank-one component.
Thus, in what follows, we analyze the operator Schmidt rank of $\hat{H}$, which immediately yields that of $\hat{G}$.

Expanding $\hat{H}$ in the local basis
$\{\hat{I}_i,\hat{Z}_i\}$ as in Eq.~\eqref{eq:operator_G}, the
coefficient tensor $g_{p_1\cdots p_N}$ has nonzero entries only when
exactly two local operators are $\hat{Z}_i$, in which case the corresponding coefficient is
$M_{ij}$. 
Here, we use the shorthand $\bm{0}_k$ to denote a length-$k$ sequence of zeros in the multi-index, so that $g_{\bm{0}_{i-1}, 1, \bm{0}_{j-i-1}, 1, \bm{0}_{N-j}} = M{ij}$.
Thus the structure of $g$ in Eq.~\eqref{eq:operator_G} is entirely determined by that of the symmetric matrix $M$.

For a given cut
$\{1,\dots,\ell\}\,|\,\{\ell+1,\dots,N\}$, we write the coupling matrix
$M$ in block form as
\begin{equation}
  M =
  \begin{pmatrix}
    M_{LL} & M_{LR} \\
    0 & M_{RR}
  \end{pmatrix},
\end{equation}
so that the QUBO operator can be decomposed as
\begin{equation}
  \hat{H} = \hat{H}_L 
          + 
          \hat{H}_R
          + \hat{H}^{(\ell)}~,
\end{equation}
where $\hat{H}_L$ and $\hat{H}_R$ contain couplings within
the left and right subsystems corresponding to upper triangular matrices $M_{LL}$ and $M_{RR}$,
and
\begin{equation}
  \hat{H}^{(\ell)}
  = \sum_{i\le \ell < j} M_{ij}\,\hat{Z}_i\hat{Z}_j
\end{equation}
collects all terms that couple the two sides of the cut.
The operator $\hat{H}^{(\ell)}$ is completely determined by the
off-diagonal block $M_{LR}$.
Taking a singular value decomposition of $M_{LR}$,
\begin{equation}
  M_{LR}
  = \sum_{\alpha=1}^{r^{(\ell)}}
      \lambda^{(\ell)}_\alpha\,
      u^{(\ell)}_\alpha v^{(\ell)\mathsf T}_\alpha,
\end{equation}
with $r^{(\ell)} = \mathrm{rank}(M_{LR})$,
we can write
\begin{equation}
  \hat{H}^{(\ell)}
  = \sum_{\alpha=1}^{r^{(\ell)}}
      \lambda^{(\ell)}_\alpha\,
      \sum_{i\le\ell} \bigl(u^{(\ell)}_\alpha\bigr)_i\,\hat{Z}_i 
      \sum_{j>\ell} \bigl(v^{(\ell)}_\alpha\bigr)_j\,\hat{Z}_j~.
\end{equation}
Thus the operator Schmidt rank of $\hat{H}$ across this cut is given by $r^{(\ell)} + 2$, where the additional two come from the rank-one contributions
$\hat{H}_L$ and $\hat{H}_R$.
Since $M_{LR}$ is an $\ell\times (N-\ell)$ matrix, we have
\begin{equation}
  r^{(\ell)} \le \min(\ell,N-\ell) \le \lfloor N/2\rfloor~.
\end{equation}
Furthermore, $r^{(\ell)} \le r$, where
$r = \mathrm{rank}(M)$ is the number of nonzero eigenvalues of the full
matrix $M$.
Combining these observations and taking the maximum over all cuts
yields
\begin{equation}
\label{eq:bound_chi}
  \min\!\left(\left\lfloor\frac{N}{2}\right\rfloor+2,r+2\right)~,
\end{equation}
which establishes the strict bound on the bond dimension of the
QUBO MPO.
Note that again this bound is invariant under adding any constant shift to $\hat{H}$ to get $\hat{G}$.

In the case of a HUBO operator, an analogous bound can be derived by considering the operator Schmidt rank of the coupling terms that cross the cut.
In practice, this can be obtained by reshaping the coefficient tensor of $\hat{H}$ into a matrix across the bipartition and taking its rank (equivalently, the number of nonzero singular values). 
This rank, together with the within-subsystem contributions, yields an upper bound on the MPO bond dimension, in direct analogy with the QUBO case.

\end{document}